# Optical Networks for Composable Data Centers

Opeyemi O. Ajibola, Taisir E. H. El-Gorashi, and Jaafar M. H. Elmirghani, *Fellow, IEEE*

*Abstract*—Composable data centers (DCs) have been proposed to enable greater efficiencies as the uptake of on-demand computing services grows. In this article we give an overview of composable DCs by discussing their enabling technologies, benefits, challenges, and research directions. We then describe a network for composable DCs that leverages optical communication technologies and components to implement a targeted design. Relative to the implementation of a generic design that requires a (high capacity) dedicated transceiver on each point-to-point link on a mesh optical fabric in a composable DC rack, the targeted design can significantly reduce capital expenditure (by up to 34 times) because fewer transceivers are used. This is achieved with little or no degradation of expected performance in composable DCs.

## I. Introduction

DIGITAL transformation in public and private organizations and the rise in household usage of digital technologies, such as IoT and cloud computing, are driving the pervasive demand for computing capacity across the globe. Consequently, the number of data centres (DCs) across the globe is increasing to accommodate the rising demands for remote computation. Amid the rapid growth in usage and the need to deploy more infrastructure, there is a strong desire for cost efficient and sustainable scaling of computing capacity for financial and environmental reasons alike.

In the last two decades, software centric approaches, such as server virtualization, containerization, micro service architecture, software defined infrastructure (SDI) and serverless computing, have been adopted in DC environments to improve efficiency [1]. However, these software-centric approaches are sub-optimal relative to the expected future growth of computing services usage. Recently, hardware and infrastructure centric measures have also been taken to complement software centric approaches. Forecasts, such as a prediction that DCs will account for 3-13% of global electricity consumption in 2030 relative to 1% in 2010 [2], are primary motivators for the development of hardware and infrastructure centric approaches. An example of such measures is the introduction of high-capacity multi-core processors with multi-threading features and lower power consumption. Hence, more computation power is enabled at lower power consumption and low physical footprint. Operators of large DCs also often construct computing infrastructure near renewable energy sources (such as wind and solar) to reduce the network and computing carbon footprint [3]. However, sole adoption of this approach can lead to performance degradation because temporal and geographical availability of renewable energy and user population distribution are not always in sync. Hence, cloud infrastructure providers are known to offset their carbon footprint via bulk purchase of renewable energy through power purchase agreements [4].

Nevertheless, the server-centric architecture deployed in conventional DCs is limited by the rigid resource utilization scope and the capacity of the server's intrinsic resources. The software centric and hardware centric approaches proposed for DCs are less effective at mitigating these challenges of conventional DCs. In recent times, the composable DC paradigm has been proposed to address the limitations of conventional DCs [5], [6]. A combination of SDI and resource disaggregation concepts in conjunction with the availability of a suitable network fabric are required to implement a composable DC. Designing a suitable network for composable DCs is a challenging task. Nonetheless, focused adoption of optical communication technologies can be leveraged to enable low cost and right-sized networks for composable DCs. This can enable greater network cost efficiency in composable DC over the adoption of a generic design. In this article, we review composable DCs by considering enabling technologies, benefits, challenges, and research directions. Secondly, we demonstrate the financial advantage of the targeted use of next-generation optical communication technologies to design composable DC network (DCN) relative to the generic adoption of such technologies in composable DC racks.

## II. Composable Data Centers Overview

Composable DCs deploy software driven techniques for on-demand composition, decomposition, and re-composition of logical computing nodes to support applications. This is achieved via the orchestration of physical or logical pools of disaggregated compute components that are inter-connected over appropriate networks.

### A. Enabling Technologies

The primary enabling technologies of composable DCs are resource disaggregation, software defined infrastructure (SDI), optical communication, and silicon photonics (SiPh).

This work was supported by the Engineering and Physical Sciences Research Council (EPSRC), in part by INTelligent Energy aware NETworks (INTERNET) under Grant EP/H040536/1, in part by SwiTching And tRansmission (STAR) under Grant EP/K016873/1, and in part by Terabit Bidirectional Multi-user Optical Wireless System (TOWS) project under Grant EP/S016570/1. All data is provided in the results section of this paper. The first author would like to acknowledge his PhD scholarship awarded by the Petroleum Technology Trust Fund (PTDF), Nigeria.

The authors are with the School of Electronic and Electrical Engineering, University of Leeds, Leeds, LS2 9JT, U.K. (e-mail: el14oa@leeds.ac.uk; t.e.h.elgorashi@leeds.ac.uk; j.m.h.elmirghani@leeds.ac.uk).

*1) Resource Disaggregation*

Resource disaggregation is essential in composable DCs. It addresses the resource stranding problem associated with conventional DCs. Resource stranding is the primary cause of the poor utilization efficiency associated with traditional servers in conventional DCs. The problem occurs because the chassis of a traditional server acts as physical utilization boundary for the server's intrinsic resources. For example, the server CPU may have spare capacity, but the I/O may be fully utilized and thus the spare CPU capacity cannot be accessed because of the rigid server "box". Hence, if the capacity of a unique resource type within a given server is unable to support additional application consolidation, components of all other unique resource types also become unavailable irrespective of each resource's idle capacity. Rack 1 in Fig. 1 shows the impact of resource stranding as input applications A-G are provisioned. Resource disaggregation mitigates this problem in DCs, where servers of the 'correct size' are composed for a given job and are disintegrated at the end of the job.

Disaggregation of server intrinsic components can be performed at different utilization scopes to achieve the intended benefits. The utilization scope of a disaggregated component refers to the maximum utilization boundary which must be maintained when components are selected to form of a logical server. The utilization scope of disaggregated components can be controlled physically or logically. Additionally, both approaches can be combined to achieve hybrid disaggregation.

**Physical Disaggregation**: Physical disaggregation entails physical separation of compute components into pools of homogenous resource types. A homogenous resourced pool can take the form of a server-like node or a sled which can fit into a standard rack-chassis. Such a node comprises of one or more compute components of the same resource type and it is the basic unit of all physically disaggregated composable DCs as shown in Fig. 1. The utilization scope of each homogeneous resourced pool can be node-limited, rack-limited, or pod-limited relative to the size of the DC under consideration. The allocation of nodes and their corresponding utilization scope determines the scale of disaggregation in composable DCs as shown in Fig. 1.

The scales of disaggregation are rack-scale, pod-scale and DC-scale [5], [6]. At rack-scale, the composable DC comprises of multiple racks, each rack comprises of multiple (homogeneous) nodes of different resource types and the utilization scope of each node in each rack is rack-limited. Rack 2 in Fig. 1 represents physical disaggregation at rack-scale. At pod-scale, the composable DC comprises of multiple racks, each rack comprises of multiple homogeneous resourced nodes and the utilization scope of each node in each rack is pod-limited as shown in Racks 5-7 of Fig. 1. Finally, at DC-scale, the composable DC comprises of multiple racks, each pod in the DC comprises of homogenous resourced racks and each rack comprises of multiple homogeneous resourced nodes.

**Logical Disaggregation**: The utilization scope of logically disaggregated resources is controlled virtually and on-demand by enforcing resource utilization boundaries via a central control entity with global knowledge of the infrastructure's state. Logical disaggregation allows the reuse of traditional server chassis [7]. Hence, a node in a logically disaggregated DC may comprise of heterogeneous compute components like traditional servers. Logical disaggregation can be applied to all variants of the physically disaggregated DC. However, the utilization scope of heterogeneous resourced nodes is no longer limited by node, rack, or pod physical boundaries. Therefore, all resource components in the DC are available for the instantiation of logical servers.

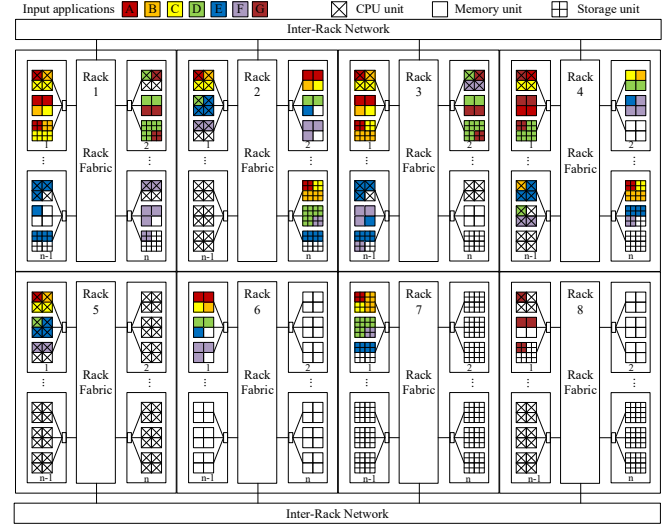

Fig. 1. Disaggregation in composable DC infrastructure.

This figure shows different forms of disaggregation in composable DCs. Rack 1 comprises of traditional servers that do not implement disaggregation. Resource components in Rack 2 are physically disaggregated at rack-scale; the rack comprises of homogeneous compute nodes. Resource components in rack 3 are logically disaggregated at rack-scale. Racks 4 and 8 implements hybrid disaggregation; 1 compute node in Rack 4 is heterogeneous while others are homogeneous. Rack 5, 6, and 7 jointly implement physical disaggregation at pod-scale.

TABLE I
INFRASTRUCTURE TEMPLATE OF APPLICATIONS

| Apps | CPU Units | RAM Units | Storage Units | Intra-logical server latency |
|---|---|---|---|---|
| A | 1 | 2 | 1 | Pod |
| B | 1 | 1 | 2 | Pod |
| C | 2 | 1 | 1 | Pod |
| D | 1 | 2 | 3 | Pod |
| E | 3 | 1 | 2 | Pod |
| F | 2 | 3 | 1 | Pod |
| G | 1 | 2 | 1 | Node |

This table gives the infrastructure template of input applications A-G. The infrastructure template of an application is the logical host configuration needed to support that application. Each infrastructure template may have CPU, memory and storage resource demands and an abstracted service level agreement (SLA) for the optimal performance of the application it supports. The maximum tolerated intra-logical server latency between the resource components of a given application represents the SLA requirement of the application.

**Hybrid Disaggregation**: A composable DC that adopts hybrid disaggregation combines both physical and logical disaggregation concepts to achieve greater efficiency by creating logical servers that support application specific requirements. For example, multiple applications to be hosted in the same composable DC may have different intra-logical server latency requirements as illustrated in Table I. The setup in Rack 4 of Fig. 1, comprising of a mix of homogenous resourced nodes and heterogeneous resourced nodes, enables a composable DC capable of supporting all input applications listed in Table I with optimal efficiency. This contrasts with the situation when physical disaggregation is performed at rack-scale or pod-scale where application G must be rejected when physical disaggregation is employed (at rack-scale or pod-scale) because of the violation of its intra-logical server latency.

*2) Software Defined Infrastructure*

A SDI is an infrastructure that can mutate on-demand to concurrently satisfy changing users' demands and its provider's desires by leveraging deep monitoring and underlying infrastructure capabilities [8]. Softwarization and automation of the underlying infrastructure's hardware resources, such as compute, storage, and network, are primary enablers of SDI. These resources are softwarized independently and become programmable. Consequently, software defined compute (SDC), software defined storage (SDS) and software defined networking (SDN) are achieved within DCs. A complete SDI is further supported by centralized controllers, deep monitoring of underlying resources and a management sub-system [8]. SDI is preceded by virtualization which abstracts resources from their underlying physical hardware. Virtualization enables the creation of logical resources from physical resources of a traditional server. These logical resources are subsequently aggregated within a single server to form logical servers that can support applications seamlessly.

In the computing domain, virtual compute resources may be instantiated as a bare metal, a VM or a container. SDC dynamically controls the instantiation, termination, and migration of virtual computing infrastructures to satisfy user demands and provider's constraints. SDS provides similar functions for virtual storage in the SDI. These virtual storage resources are created from abstracted heterogeneous storage components which are controlled centrally. The logical resources of a virtual infrastructure are connected by virtual networks which are overlaid on underlying physical networks to satisfy the SLAs of logical servers. Furthermore, virtual networks in SDIs also provide communication paths between logical servers and between the logical servers and other external systems. SDN implements dynamic mutation of virtual networks to ensure continual satisfaction of desired objectives in the SDI. Composable DCs leverage and extend features of SDI in conjunction with resource disaggregation to achieve greater efficiency [7]. After the adoption of disaggregation to separate compute components into pools of data-plane components, SDI enables a control-plane that remotely manages and orchestrates the underlying compute components in each pool to provision on-demand logical servers that support applications and services.

*3) Optical Communication Technologies*

Disaggregation of computing resources in composable DCs brings about the need for a high bandwidth, low latency, flexible and energy efficient interconnection between logically and/or physically separated compute components. The many benefits of optical networking technologies (such as low latency, low interference, long-distance, high bandwidth, low energy, and scalable communications) make them better candidates compared to electrical networking technologies for such a network. However, buffering in the optical switching domain remains challenging, especially in electronic-based computing infrastructures where network congestion occurs frequently. Furthermore, historically, optical network components often have a large footprint, and their manufacturing processes are generally more expensive relative to the cheap and matured manufacturing process of silicon micro-electronics circuits used in electronic network technologies.

The advent of SiPh technologies enabled the emergence of affordable hybrid networks designed using novel opto-electronic circuits. These circuits, which integrate electronics and photonics into a single circuit, are manufactured using cost effective and matured traditional silicon IC manufacturing processes [9]. Hence, opto-electronic networks that combine features of both electronic and optical network technologies are enabled. The resulting hybrid network addresses some challenges of electronic network technologies and avoids challenges of optical network technologies while leveraging the advantages of both. This is achieved through integrated SiPh components manufactured as photonic integrated circuits that can effectively meet capacity, flexibility, energy efficiency, and scalability requirements in composable DCNs.

SiPh technology is widely used to manufacture optical transceivers adopted in DCs and data communication networks. Attractive features of SiPh (such as dense integration, higher data rates, long reach, low energy per bit (few pJ/bit) and low manufacturing cost) makes it a preferred choice for composable DCs. Forecasted drop in the cost of SiPh-based interconnects to $1/Gbps is also expected to further encourage the adoption of the technology [10]. The emergence of densely integrated SiPh switches is expected to support the commercial implementation of composable DCs. For instance, densely integrated hybrid packet switching devices, which co-package optics and ASIC switch chips, may be adopted to implement switches with optical IO, to design high-speed coherent integrated circuits or to manufacture system on chip (SoC) with integrated fabric switch [11] in composable DCs. SiPh may also be adopted in the on-board DCN tier of composable DCs; hence, enabling high speed and power efficient optical IO between switch chips and NICs and between compute components within nodes of composable DCs.

*B. Benefits of Composable DC*

The benefits of composable DCs include increased modularity, greater agility and flexibility and improved efficiency.

*1) Modularity*

The shift from server-wise resource utilization to component-wise resource utilization in composable DCs enables increased modularity and promotes proportional usage of underlying compute components. Furthermore, the modular design of underlying compute components in composable DC also enables precise upgrade and replacement of compute components.

*2) Agility and Flexibility*

Intelligent control and deep monitoring of the underlying (hardware) compute components of composable DCs, as enabled by SDI, supports greater agility and flexibility. The ability to guide automated transformation of the infrastructure on-demand with insights derived from intelligent analysis of collected monitoring data is responsible for this. On-demand slicing and aggregation of physical and virtual compute components in composable DC enables an adaptable infrastructure that can scale dynamically at run-time based on temporal resource-demands of applications. Applications can also be migrated on-demand to ensure optimal performance via the automation of the composable DC.

*3) Greater Efficiencies*

Composable DCs enable greater efficiency in multiple domains. Component-wise resource utilization in composable DCs enables proportional usage of resource capacity which leads to higher resource utilization efficiency relative to the conventional DC. Therefore, the number of active resource components and total power consumed in composable DC are lower [7]. This leads to improved energy efficiency because only necessary components are turned on while other components remain in an inactive state. Thus, the operational expenditure (OPEX) of infrastructure providers is curtailed. The number of required resource component in a composable DC is relatively lower than that of the conventional DC due to increased utilization of active resource components. Hence, capital expenditure (CAPEX) is reduced because fewer components are purchased, and the footprint of primary and auxiliary infrastructures is also reduced. Increase in the precision of upgrading and replacing modular resource components of composable DC can lead to further CAPEX reductions.

The adoption of software and hence programmability to control the underlying hardware components on-demand enables service-oriented consumption of computing capacity which can support dynamical repurposing of idle compute capacities to provision other applications. This also promotes greater efficiency and generates additional revenues that can help infrastructure providers to offset their total cost of ownership. Adoption of in situ data processing in composable DC also enables greater network efficiency in instances where in-memory communication or data sharing is required. This is because data is processed at its location; hence, transmission over networks is minimized.

*C. Implementation Challenges and Research Directions*

Relative to conventional DC, the concept of composable DC introduces a range of implementation challenges. Such challenges span domains such as physical networks, SDI, high availability and fault tolerance and application development and design.

*1) Physical Networks*

Exposure of essential intra-logical server communications onto higher tiers of DCNs introduces several challenges that may constrain the full implementation of the composable DC. These challenges are primarily associated with the CPU-memory and CPU-CPU communications which require ultra-high communication bandwidth and ultra-low latency for efficient performance of some applications. The physical limitations of traditional network media (such as fibre optics and copper media) control the maximum practical distance between disaggregated components which form a logical host. However, shorter communication distances in DC environments relative to the high speed and low attenuation of light propagation in optical fibre implies that propagation delay has minimal impact when optical media are adopted. Although, the adoption of optical communication provides a solution to many network challenges of composable DCs, today's optical technologies are somewhat limited [6]. For instance, if optical DC interconnects must stay competitive against electronic DC interconnects, the energy efficiency values stated in [12] should be achieved at the different hierarchies of DC interconnects. Furthermore, the traditional network protocols and software stack [11] also introduce additional delays on the communication paths between source and destination resource-components.

Parallel high-speed optical communication paths can be adopted to ensure appropriate capacity for communication between physically disaggregated computing components that form logical servers. However, integration of multiple ports and switches onto computing components in such a scenario introduces additional challenges. Adoption of SiPh technologies promises potential solutions to these challenges; however, as the number of disaggregated components increases in composable DCs, the design of scalable physical network topologies must consider criteria such as flexibility, cost, power consumption and port/interface count to find practical solutions.

The traditional network software stack can be streamlined to minimize the additional delays introduced by the stack. However, this may be sub-optimal to satisfy the ultra-low latency communication between CPU and memory/CPU as required by some applications. A complementary solution could use the network software stack to relax inter-resource latency requirements by adopting techniques that amortize the latency related performance penalties in composable DCs. Furthermore, revisions in component hardware architecture may also be adopted to mitigate network challenges introduced by disaggregation. For instance, the use of intermediate hardware buffers or memory/data stores could be standardized in the physical architecture of composable DCs as observed in some prototypes of partially disaggregated DCs [13], where primary and secondary memory tiers are implemented to minimize performance degradation.

Another network challenge which emerges with the advent of composable DC is the heterogeneity of interconnect interfaces between disaggregated resources. The heterogeneity of interfaces is expected to increase as the number, types and versions of components increases in composable DCs. The proposition of a universal interface for components in composable DC by the Gen-Z consortium [14] is a step in the right direction to address this problem. Amidst several physical challenges of the networks of composable DC, exploring techniques to achieve greater flexibility and scalability over physical network topology of the composable DC is also important. This is crucial in situations where slicing, aggregation and sharing of physical components is required.

*2) Software Defined Infrastructure*

Orchestration and management software in composable DC must support on-demand slicing and aggregation of underlying resource components to satisfy user demands. Relative to the SDI deployed in conventional DCs, this introduces additional management, operation, and maintenance complexities and overheads in composable DCs. The use of additional tiers of abstraction in the management and control system of composable DCs may reduce such complexities. Furthermore, control plane elements in composable DCs can implement policies that ensure optimal functioning of the infrastructure. An example of such policies may ensure optimal placement of applications resource demands in composable DCs in real-time. Solving similar problems in conventional DCs is difficult. This difficulty grows in composable DCs because of increased modularity.

*3) High Availability*

The difficulty of achieving high availability (HA) increases in composable DCs relative to conventional DCs. This is because HA must be implemented in both logical and physical layers of the data-plane (hardware) to ensure reliability and fault tolerance. For example, the process of creating a back-up (logical) server requires components that are physically disjoint from components that form the primary (logical) server to create a fail-safe system. Hence, the management and orchestration software of composable DC must support redundancy across physical and logical hardware domains.

*4) Application Design and Development*

The emergence of composable DCs motivates re-evaluation of the application design and development process to study the impact of the concept on the performance and suitability of legacy, present and future applications. Legacy applications may fail to run on composable DCs if the infrastructure design requires the optimization of applications for the novel computing platform. On the other hand, the optimal efficiency of composable DC may be reduced if the details of disaggregation are over-abstracted from the applications running on the platform [1]. Hence, the right balance must be struck between applications design and development processes and infrastructure details abstraction.

## III. TARGETED NETWORK FOR COMPOSABLE DCS

In [7], [15], we proposed two variants of a targeted design that leveraged optical technologies to enable a suitable network for composable DCs. The targeted design employed optical components, wavelength division multiplexing (WDM) and space division multiplexing (SDM) and higher single lane data rate to achieve comparable performance as a generic design while maintaining full mesh connectivity within the rack and utilizing fewer SiPh transceivers.

*A. Network Design Description*

Optical components employed by the targeted network design include SiPh transceivers, semiconductor optical amplifier (SOA)-based switches, circulators, de-multiplexers, combiners, and a mesh optical backplanes within each rack. Each node can have two interfaces as illustrated in Fig. 2; each interface comprises of a set of SiPh transceivers. The number of transceivers at each interface is determined by the maximum throughput required of each implementation of the targeted network design. The set of wavelengths transmitted and received at an interface must be mutually exclusive. Furthermore, both interfaces of a compute node must transmit and receive disjoint sets of wavelengths. These conditions are satisfied by ensuring that the first interface transmits the set of wavelengths that is received by the second interface and that the second interface transmits the set of wavelengths that is received by the first interface. This promotes seamless replication because each compute node in a rack of a composable DC implements the proposed network design in Fig. 2. It also promotes the benefits of economies of scale in a large deployment.

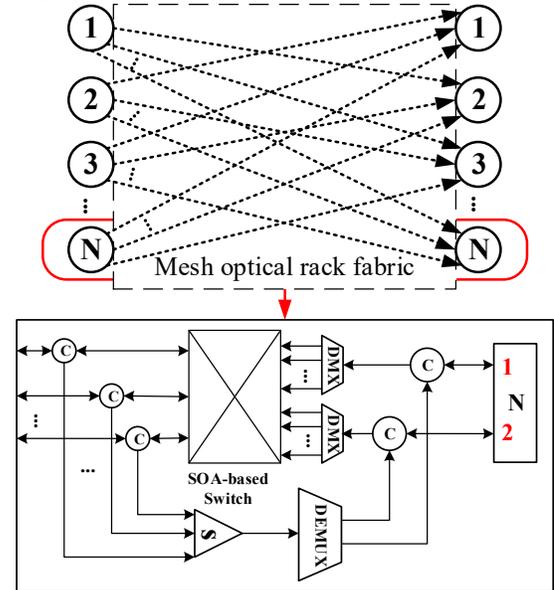

Fig. 2. Proposed network for composable DC rack with node interface design.

Optical switches in the design ensure that a wavelength is only transmitted to an intended destination node via the low latency optical mesh backplane while combiners receive the ingress traffic (on the selected wavelengths destined for each node) from the optical mesh backplane. Hence, optical switches and combiners collectively ensure that optimal (fewer) number of SiPh transceivers are required for each node to communicate over the optical backplane in each rack. The proposed network leverages the fact that concurrent all-to-all communication is not mandatory between compute nodes in the same rack of composable DCs. Hence, each compute node can effectively and efficiently utilize its limited capacity for targeted communication instead of implementing a generic design that leads to significant wastage. Note that an implementation of a generic design requires a dedicated SiPh transceiver for each point-to-point link on the optical backplane illustrated in Fig. 2. Full description of the two variants of the proposed targeted design and their corresponding operation principle are given in

[15]. Additionally, mixed integer linear programming (MILP) optimization models are also presented in [15] to represent the proposed variants of the targeted design and to demonstrate the efficacy of the design in a rack-scale composable DC that implements physical, logical and hybrid disaggregation. In this article we focus solely on the cost performance evaluation of the proposed targeted design relative to a generic design.

*B. Cost Performance Evaluation*

We solved the MILP optimization model proposed in [7], [15] for the targeted design to obtain the maximum network throughput. A single rack comprising of different numbers of compute nodes is considered. It is assumed that each interface of a compute node in the targeted design has 4 transceivers that transmit and receive a single wavelength each at the single wavelength data rate of 100 Gbps. Hence, each compute node in the targeted design has a maximum capacity of 800 Gbps. On the other hand, the maximum capacity of each transceiver deployed on each point-to-point link on the optical backplane in the generic design is 800 Gbps. In all instances, the cost efficiency of integrated SiPh interconnects used in the generic design is assumed to be $1/Gbps as expected in next generation networks [10]. On the other hand, the cost efficiency of integrated SiPh interconnects used in the targeted design is varied between $1/Gbps and $40/Gbps. Fig. 3 shows the resulting CAPEX of both generic and our targeted design under different numbers of compute nodes per rack. Note that, the maximum throughput under both designs is easily inferred when the cost efficiency of integrated SiPh interconnects is $1/Gbps as shown in Fig. 3. As expected, the generic design achieved significantly higher throughput than the targeted design.

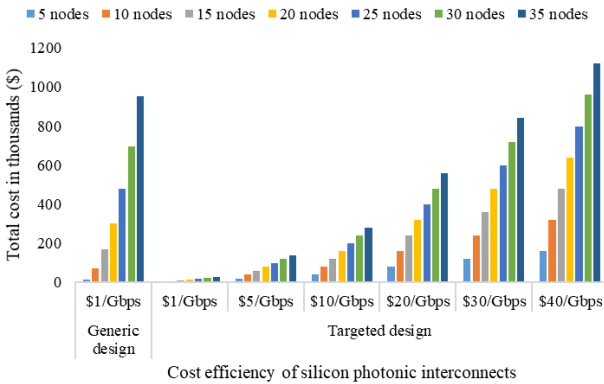

Fig. 3. Total cost of generic and targeted design approaches

However, despite reduced network throughput when the targeted design is implemented, the results in [15] showed that relative to the adoption the generic design, composable DC performance is minimally impacted. At equal cost efficiency of $1/Gbps, Fig. 3 shows significantly lower cost when the targeted design is implemented instead of the generic design. Furthermore, Fig. 3 and Fig. 4 show that relative to the generic design, the performance of targeted design improves as the number of compute nodes in each single rack increases. The trend in Fig. 3 and Fig. 4 is that the cost of the targeted design is $N-1$ times lower than that of the generic design; where, $N$ is the number of nodes in the rack. For instance, when a rack with 35 compute nodes is considered, at cost efficiency of $1/Gbps for both designs, the targeted design is 34 times more cost efficient than the generic design as shown in Fig. 4.

Note that, the cost ratio in Fig. 4 is the ratio of the total network cost of a generic design at $1/Gbps to the total network cost of a targeted design at $Y$/Gbps; where, $Y$ is the corresponding cost efficiency of SiPh interconnects used in the targeted design. The cost ratio is directly proportional to the number of nodes in the rack and inversely proportional to the cost efficiency of SiPh interconnects used in the targeted design. Hence, as the number of compute nodes in the rack increases, a cost ratio that is above 1 can be sustained provided that the cost of SiPh interconnect used in the targeted design is only moderately higher than those used in the generic design.

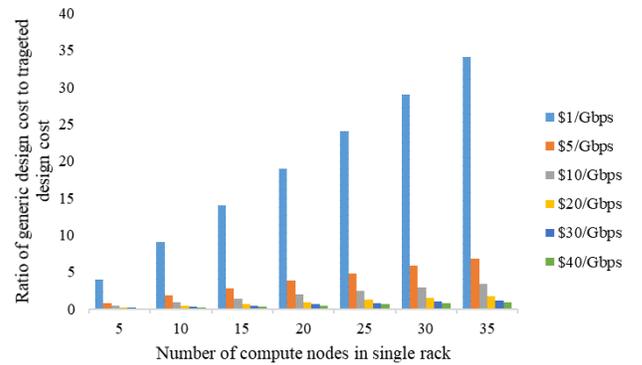

Fig. 4. Ratio of generic design cost to targeted design cost

The results in Fig. 4 show that at a lower number of compute nodes per rack, the cost efficiency of the targeted design over the generic design rapidly disappears as the cost efficiency of SiPh interconnects used in the targeted design reduces. However, as the number of compute nodes per rack approaches those expected in a realistic DC, a significant drop in the cost efficiency of SiPh interconnects adopted by the targeted design is required for the generic design to achieve greater cost efficiency. For instance, in a rack with 35 compute nodes, the cost efficiency of SiPh interconnects used in the generic design must be 40 times greater than those used in the targeted design to obtain a cost ratio that is below 1. Such scenario is not expected in practice because commensurate improvements in the cost efficiency of SiPh interconnects used in the generic design are expected for those used in targeted designs as well. Hence, relative to the generic design, the greater cost efficiency of the targeted design will persist in a realistic composable DC with multiple compute nodes per rack when the cost of SiPh interconnects used in targeted design is equal or moderately higher than those used in the generic design.

## IV. CONCLUSIONS

In this article, we reviewed composable DCs by highlighting their enabling concepts, benefits and implementation challenges and research directions. Additionally, we reviewed a network topology for composable DCs that implements targeted use of optical communication technologies and components while maintaining mesh connectivity between co-rack compute nodes in a composable DC. Relative to the implementation of generic design over a mesh optical rack backplane, the targeted design uses fewer transceivers with little or no performance degradation. Compared to the generic design, the targeted design significantly reduced the CAPEX in realistic composable DC with high number of compute nodes per rack.

**Opeyemi O. Ajibola** is working towards the PhD degree in the School of Electronic and Electrical Engineering, University of Leeds, Leeds, UK.

**Taisir E. H. El-Gorashi** is currently a Lecturer in optical networks in the School of Electronic and Electrical Engineering, University of Leeds. Previously, she held a Postdoctoral Research post at the University of Leeds (2010–2014), where she focused on the energy efficiency of optical networks investigating the use of renewable energy in core networks, green IP over WDM networks with datacenters, energy efficient physical topology design, energy efficiency of content distribution networks, distributed cloud computing, network virtualization and big data.

**Jaafar Elmirghani** is Fellow of IEEE, Fellow of IET and Fellow of Institute of Physics. He was PI of the £6m EPSRC Intelligent Energy Aware Networks (INTERNET) Programme Grant. He is Co-Chair of the IEEE Sustainable ICT initiative, a pan IEEE Societies initiative responsible for Green ICT activities across IEEE, 2012-present. He was awarded in international competition the IEEE Comsoc 2005 Hal Sobol award, 3 IEEE Comsoc outstanding technical achievement and service awards (2009, 2015, 2020), the 2015 GreenTouch 1000x award, IET Optoelectronics 2016 Premium Award for work on energy efficiency and shared the 2016 Edison Award in the collective disruption category with a team of 6 from GreenTouch for joint work on the GreenMeter. His work led to 5 IEEE standards with a focus on cloud and fog computing and energy efficiency, where he currently heads the work group responsible for IEEE P1925.1, IEEE P1926.1, IEEE P1927.1, IEEE P1928.1 and IEEE P1929.1 standards. He has published over 550 technical papers, and has research interests in energy efficiency, optimization, cloud and fog computing.